\documentclass[hidelinks]{article}

\usepackage{arxivstyle}

\usepackage[utf8]{inputenc} 
\usepackage[T1]{fontenc}    
\usepackage{hyperref}       
\usepackage{url}            
\usepackage{booktabs}       
\usepackage{amsfonts}       
\usepackage{nicefrac}       
\usepackage{microtype}      
\usepackage{float}
\usepackage{amsmath}
\usepackage{amssymb}
\usepackage{authblk}
\usepackage{fancyhdr}       
\usepackage{graphicx}       

\begin{document}

\title{Comparison of saturation rules used for gyrokinetic quasilinear transport modeling}

\author[1,2]{Scott E. Parker}
\author[2]{Calder Haubrich}
\author[1]{Qiheng Cai}
\author[2]{Stefan Tirkas}
\author[3]{Yang Chen}
\affil[1]{Renewable and Sustainable Energy Institute, University of Colorado, Boulder}
\affil[2]{Department of Physics, University of Colorado, Boulder}
\affil[3]{Center for Integrated Plasma Studies, University of Colorado, Boulder}

\date{\today}

\maketitle

\begin{abstract}
Theory-based transport modeling has been widely successful and is built on the foundations of quasilinear theory.  Specifically, the quasilinear expression of the flux can be used in combination with a saturation rule for the toroidal mode amplitude. Most transport models follow this approach.  Saturation rules are heuristic and difficult to rigorously derive. We compare three common saturation rules using a fairly accurate quasilinear expression for the fluxes computed using local linear gyrokinetic simulation. We take plasma parameters from experimental H-mode profiles and magnetic equilibrium and include electrons, Deuterium, and Carbon species. 
We find that the various saturation rules give qualitatively similar behavior. This may help explain why the different theory-based transport models can all predict core tokamak profiles reasonably well.  Comparisons with nonlinear local and global gyrokinetic simulations are also discussed.
\end{abstract}



\section{Introduction}
Prediction of turbulent particle and energy transport is critical for improving the performance of a fusion reactor. Much progress has been made with reduced models in the core region, from the pedestal top inwards. 
Theory-based models, including the trapped gyro-Landau fluid (TGLF) model\cite{staebler:2005,staebler:2007,kinsey:2015,staebler:2016}, the multi-mode model (MMM)\cite{kinsey:1995,bateman:1998, rafiq:2013, weiland:2020}, and the gyrokinetic transport model QuaLiKiz\cite{bourdelle:2007,bourdelle:2016,citrin:2017,stephens:2021} are successful in predicting core density and temperature profiles over a range of tokamak plasma operating conditions. Additionally, quasilinear theory is used widely to compare with both experiment and nonlinear gyrokinetic simulation\cite{dannert:2005,fable:2010,lapillonne:2011,tirkas:2023}. Typically one takes the quasilinear expression for the flux and invokes a heuristic saturation rule to obtain the mode amplitude thereby determining the nonlinear flux. While the level of the fluxes obtained using this type of approach may not be accurate, the parametric dependence on wavelength and plasma parameters is often insightful. 
 Even with the successes of the various theory-based transport models for predicting core density and temperature profiles, there is still a need for better understanding. For example, particle transport and associated density build-up is less well understood\cite{angioni:2009c,howard:2021b}. Additionally, High-Z impurities, e.g. Tungsten in ITER, will not fully ionize and can produce significant radiative power loss if core concentrations are not well controlled\cite{loarte:2020,angioni:2017}.

 Here, we further examine the quasilinear transport modeling approach. We compare to local and global nonlinear gyrokinetic simulation which best models the governing equations with relatively few approximations. We will directly compare three widely used saturation rules, two of which  come from simple scaling arguments\cite{fable:2010,bourdelle:2007}, and a third which has been shown to give reasonable parameter dependence for fluxes\cite{kumar:2021}.  While the comparisons we present are rudimentary, we are unaware of such a study of the sensitivity to the saturation rule.  We will also compare with the TGLF model which is specifically designed to agree with flux-tube gyrokinetic simulation from the CGYRO code\cite{staebler:2021,candy:2016}.  The goal of this paper is to examine the sensitivity of the choice of saturation rule which is the part of the theory least well understood.  Generally, tokamaks operate within regimes that have relatively good confinement and hence it is not unreasonable to assume the turbulence is weak and made up of a number of active linear eigenmodes that are interacting due to weak nonlinear coupling.  Linear calculations with gyrokinetic codes are routine and fast computationally. One can easily obtain linear fluxes from gyrokinetic simulation.  Nonlinear simulations are much more compute-intensive.  However, no information on the saturation level is available from linear calculations.  Therefore, it is common to invoke a "saturation rule" that gives the saturation level of the turbulence as a function of the linear growth rate, wave number, and other parameters. The capability to derive a rigorous saturation rule is elusive. One reasonable approach is to obtain an empirical saturation rule using the scaling of nonlinear gyrokinetic simulation\cite{staebler:2007}. While the saturation rule is probably the weakest link with regard to rigor, the assumption of a quasilinear expression for the flux and how the quasilinear flux is calculated may also be approximate.

  We will discuss results from gyrokinetic simulation using the GENE and GEM codes.  The GENE code will be used for linear and nonlinear local calculations, including the calculation of the quasilinear expression for the flux\cite{jenko:2000b,gorler:2011}. We use GENE for linear calculations due to its high accuracy, good convergence properties, and comprehensive physics capability. GEM is an efficient tool for nonlinear global simulation due to both its robust behavior over a wide range of parameters and fast performance on parallel computing platforms\cite{chen:2003b,chen:2007}. For this study, we choose realistic plasma profiles and magnetic equilibrium from a conventional ELMy H-mode DIII-D case (162940) just prior to the onset of an ELM\cite{hassan:2022}. We include electrons and two ion species, namely, Deuterium (main) and Carbon (impurity). Details will be discussed in Sec.~2.
We begin by discussing the plasma parameters for our study in Sec.~2. In Sec.~3, we investigate the linear properties of the selected profile. In Sec.~4, the quasilinear theory is described and the resulting turbulent transport is compared for three different saturation rules. In Sec.~5 we compare to nonlinear fluxes from the GENE and GEM codes.


\section{Tokamak plasma parameters and assumptions}
\label{sec:headings}

For comparing the three saturation rules in quasilinear theory, we use DIII-D discharge 162940 which is an ELMy H-mode case.  Magnetic equilibrium and profiles are constructed prior to ELM onset. This particular case has been used recently for electron-temperature-gradient and micro-tearing mode studies\cite{hassan:2021, hassan:2022,curie:2022}. Details about this particular case can be found in Ref.~\cite{hassan:2022}.  We use a Miller equilibrium\cite{miller:1998} and obtain the Miller parameters from EFIT equilibrium with a  $513\times513$ $(R, Z)$ grid, and density and temperature profiles.  

The purpose of this work is to directly compare theoretical models and not predict experimental transport levels. We will not include the effect of equilibrium shear flow because simple quasi-linear theory does not take this into account. Including zonal flow and cross-coupling between electron and ion scales continue to be an active research topic\cite{staebler:2016,citrin:2017}. Neglecting shear flow in the quasilinear theory will allow for a more transparent comparison of the various saturation rules.  We include realistic collisionality in the linear analysis as well as in the nonlinear gyrokinetic simulations.  Gyrokinetic ions and drift-kinetic electrons with electromagnetic fluctuations perpendicular to B ($\delta B_\perp$) will be used in the linear and nonlinear gyrokinetic simulations. $\delta B_\parallel$ will be neglected. The plasma $\beta$ is reduced for some nonlinear simulations presented in Sec.~5, and details will be discussed there.

\begin{figure*}
\centering
\includegraphics[width=0.8\textwidth]{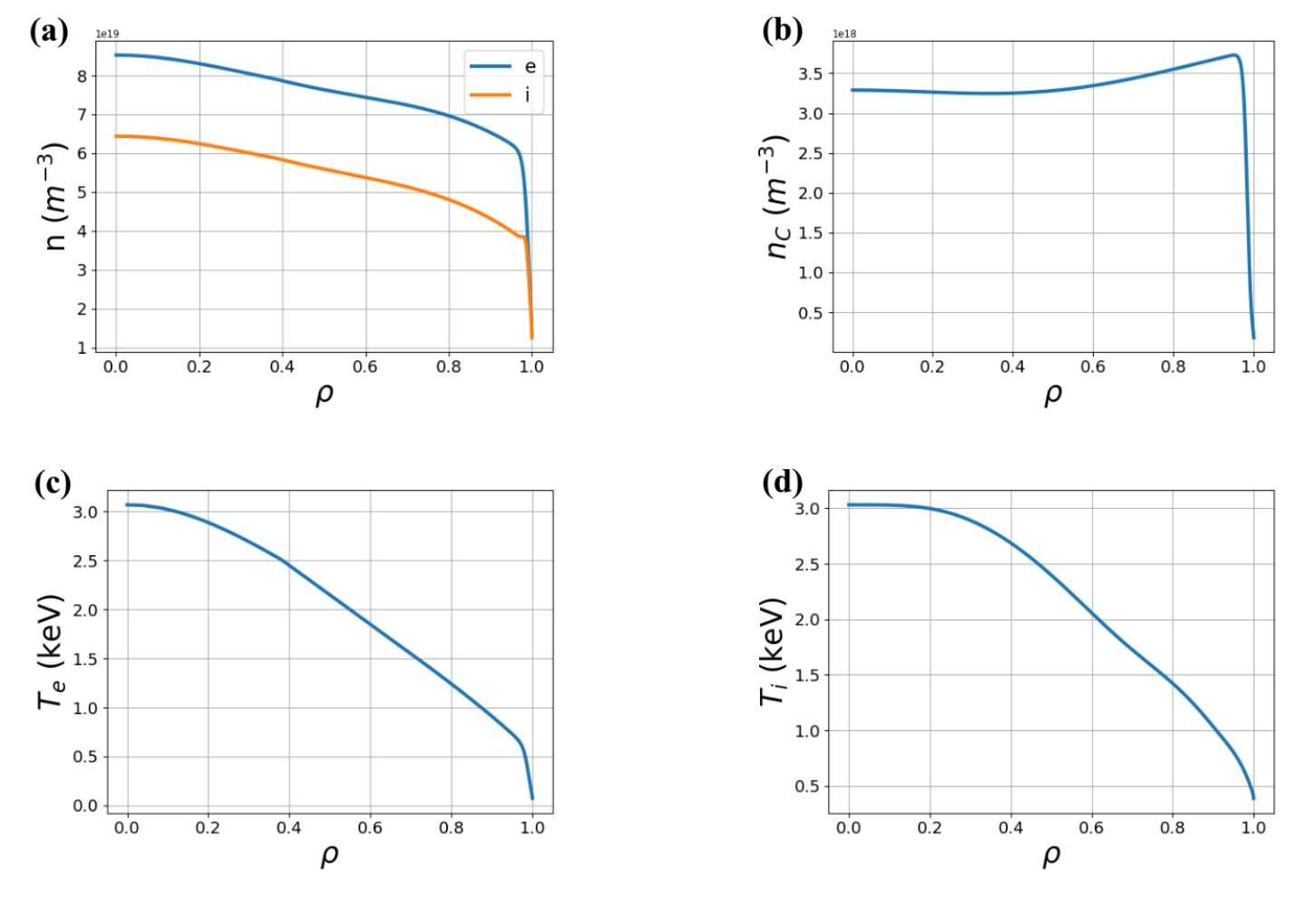}
\label{fig:fig1} 
\caption{Profiles for DIII-D 162940 ELMy H-mode just prior to ELM onset. (\textbf{a}) Electron and main ion density profiles. (\textbf{b}) Impurity density profiles. (\textbf{c}) Electron temperature profile. (\textbf{d}) Main ion temperature profile.}
\end{figure*}

Fig.~1 (a) shows the main ion (Deuterium) and electron density profiles for DIII-D 162940. (b) shows the carbon impurity density profile. (c) shows the electron temperature profile, and (d) shows the main ion temperature profile.  For this study, the impurity temperature is assumed to be equal to the main ion temperature.  And, we will only account for one impurity species, namely Carbon. The Carbon profile is hollow hence we expect inward radial flux for the impurity species.
We choose three radial locations shown in Table~I for our study, where $\rho=r/a$ and r is the Miller radial coordinate\cite{miller:1998},
\begin{equation}
r= \frac{R_{\text{max}} - R_{\text{min}}}{2},
\end{equation}
where $R_{\text{max}}$ and $R_{\text{min}}$ are the maximum and minimum major radius of each flux surface, respectively. $a$ is the value of r, from Eq.~(1), at the separatrix. 
\begin{table}[H]
\caption{Local tokamak plasma parameters at $\rho$=0.8, 0.85 and 0.9.}\label{tab1}
\centering
\footnotesize{
\begin{tabular}{|c|c|c|c|c|c|c|c|c|c|c|c|c|c|}
    \hline
    $\rho$ & $\frac{R}{L_{T_{i}}}$ & $\frac{R}{L_{T_{e}}}$ & $\frac{T_{e}}{T_{i}}$ & $\frac{R}{L_{ne}}$ & $\frac{R}{L_{ni}}$ & $\frac{R}{L_{nC}}$ & $\frac{n_{C}}{n_{e}}[\%]$ & $q$ & $\hat{s}$ & $\beta_{e}[\%]$ & $\kappa$ & $\delta$ & $\zeta$\\
    \hline
    0.8 & 6.71 & 7.49 & 0.87 & 1.44 & 2.40 & -0.71 & 5.16 & 2.28 & 1.75 & 1.00 & 1.47 & 0.21 & -0.03 \\
    \hline
    0.85 & 9.16 & 8.94 & 0.87 & 1.84 & 3.06 & -0.75 & 5.37 & 2.56 & 2.17 & 0.85 & 1.51 & 0.24 & -0.04\\
    \hline
    0.9 & 12.49 & 11.17 & 0.88 & 2.36 & 3.98 & -0.79 & 5.65 & 2.97 & 2.94 & 0.67 & 1.55 & 0.28 & -0.05\\
    \hline
\end{tabular}
}
\end{table}
Table~1 gives the local parameters at the three radial locations ($\rho$) for our analysis. We begin by examining linear stability at these three radial locations.  We note that the quasilinear analysis in Sec.~4 is a local analysis and is based on the local parameters given in Table~1.  Physical quantities such as the major radius, $B$, $n$ and $T$ are important for determining collisionality and for conversion to physical (SI) units.
$\frac{R}{L_{Ti}}$ and $\frac{R}{L_{Te}}$ are the normalized temperature gradient of ions and electrons. We assume the Carbon temperature (and temperature profile) is the same as the main ions. $\frac{R}{L_{ni}}$, $\frac{R}{L_{ne}}$ and $\frac{R}{L_{nc}}$ are  normalized density gradient of ions, electrons, and carbon. The ratio of electron temperature and ion temperature and the impurity concentration (carbon density to electron density) are given by $\frac{T_{e}}{T_{i}}$ and $\frac{n_{C}}{n_{e}}$. 
$q$ is the safety factor and $\hat{s}= {\rho \over q} { dq \over d\rho}$ is the magnetic shear parameter, and $\beta_e = \mu_0 n_e T_i / B^2$. The Miller parameters\cite{miller:1998} for elongation, triangularity and squareness are $\kappa$, $\delta$ and $\zeta$.


\section{Linear analysis}

We begin by studying the local linear properties of the tokamak plasma parameters (162940) discussed above in Sec.~2, near the pedestal top, and scan $\rho=0.8,0.85,0.9$.  We do initial-value calculations with the GENE code in the flux-tube limit.
In Fig.~2, we show the linear growth rate and real frequency for the three radial locations specified in Table~1 versus $k_y$, where $y$ is the binormal perpendicular coordinate.  In the following section, we will also use the linear output from GENE in the form of the particle and energy fluxes and the electrostatic potential linear mode structure to parameterize $k_\perp$.
\begin{figure}[h]
\centering
\includegraphics[width=0.6\textwidth]{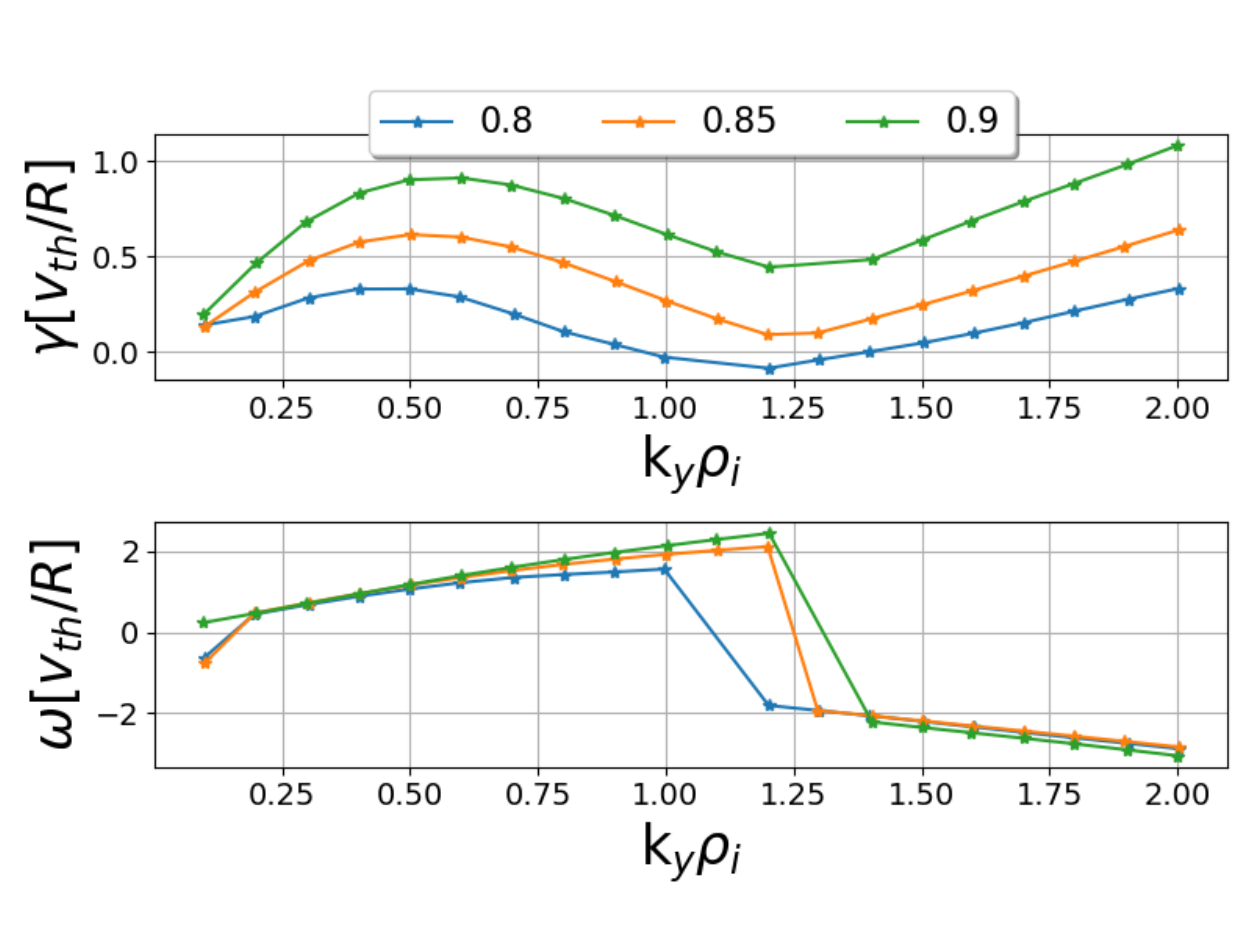}
\label{fig:fig2} 
\caption{Growth rate and real frequency at the radial locations $\rho=$ 0.8, 0.85 and 0.9 in Miller geometry from local GENE linear initial-value simulation.}
\end{figure}
 $\rho_i$ is the Deuterium species ion gyroradius and $v_{th} = \sqrt{T_i/m_i}$
The ``$i$'' subscript will refer to the main ion Deuterium species throughout the paper.
Fig.~2 has qualitative features common to core H-mode plasmas and even the so-called ``Cyclone base case''\cite{gorler:2016}. An ion mode, or ion temperature gradient (ITG) mode, for $k_y \rho_i \lesssim1.4$, and an electron mode, or a collisionless trapped-electron mode (CTEM) dominates for $k_y \rho_i \gtrsim1.4$. As one approaches the steep gradient region ($\rho=$0.85 and 0.9) in the pedestal, a negative frequency unstable mode appears at longest resolved wavelength.  This is the micro-tearing mode (MTM) which is often the dominant mode in the pedestal region for these parameters\cite{hassan:2022}. Global analysis is required to accurately model the long-wavelength (MTM).

\begin{figure}
\centering
\includegraphics[width=0.6\textwidth]{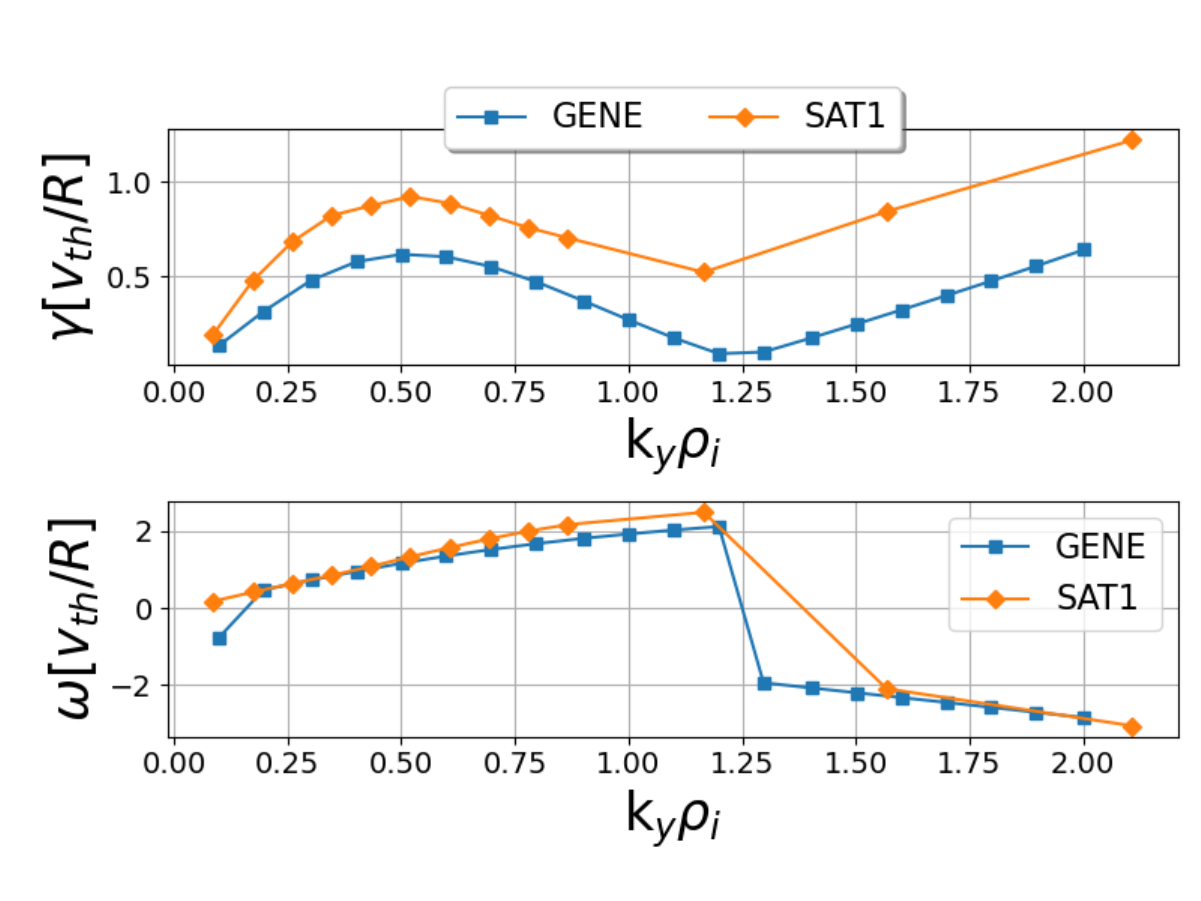}
\label{fig:fig3} 
\caption{Comparison of linear frequency and growth rate for GENE and TGLF at $\rho = 0.85$.}

\end{figure}

Fig.~3 shows a linear comparison between TGLF and local GENE at $\rho=$0.85. In Sec.~4, we will compare quasilinear theory results to TGLF as well.  The real frequencies agree very well between GENE and TGLF. We note that TGLF using the SAT1 model\cite{staebler:2016} predicts higher growth rates than GENE.  GENE is a more accurate local linear gyrokinetic calculation, but it could be that the higher growth rate is compensated by the TGLF SAT1 saturation rule so that TGLF still gives accurate fluxes. We note that global effects would typically be stabilizing, so this is another effect important for realistic modeling not accounted for here.

\section{Quasilinear theory}
\subsection{Quasilinear expression of fluxes using linear gyrokinetic simulation}

In good confinement regimes, core tokamak turbulence fluctuations are small. It is not unreasonable to assume a superposition of a finite number of linear eigenmodes at small amplitude, leading to the validity of the quasilinear expression for the fluxes. The quasilinear flux is quadratic in the mode amplitude.  What is more uncertain (or unknown) is the fluctuation amplitude, and we will discuss three plausible saturation rules in the following section.  Linear flux-tube gyrokinetic simulation is used to predict the quasilinear fluxes assuming a saturation rule.  We follow Lapillonne\cite{lapillonne:2011} prescription and use the GENE code to obtain linear fluxes.  GENE uses field line following coordinates $(x,y,z)$ where $x$ is a radial coordinate, $y$ is the other binormal coordinate and $z$ is the coordinate along the field line.  We decompose the fluxes in $k_y$ and define a general quasi-linear flux quantity $F^{ql}$, where $F$ can represent particle flux $\Gamma_\alpha$ or energy flux $Q_\alpha$ for species $\alpha$.  The linear flux is proportional to the square of the amplitude
 \begin{equation}
    F^{lin}_{k_{y}}=\hat{G}_{k_{y}} \left| \hat{\Phi}_{0,k_{y}}(z=0) \right|^{2},
\end{equation}
where we assume that the mode amplitude can be parameterized at $z=0$ (or $\theta=0$).  It is straightforward to calculate the amplitude normalized linear flux $\hat{G}_{k_{y}}$ from linear gyrokinetic simulations.  Given a saturation rule for the amplitude, the fluxes can then be calculated using
\begin{equation}
    F^{ql}=\sum_{k_{y}} A^2(k_{y})  \hat{G}_{k_{y}}^{ql}\Delta k_{y},
\end{equation}
where $\Delta k_{y}$ is the $k_y$ spacing, and $A^2(k_{y})$ parameterizes the mode amplitude and will be determined using three simple saturation rules discussed below. The saturation rule used in Lapillonne\cite{lapillonne:2011} is
\begin{equation}
    A^2(k_{y})=A^2_{0}\left( \frac{\gamma_{k_{y}}}{\left\langle k_{\bot}^{2} \right\rangle} \right)^{2}.
\end{equation}
where $\gamma_{k_{y}}$ is the linear growth rate. Eq.~(4) is one of the saturation rules we will examine. Some care is taken in determining $\left\langle k_{\bot}^{2} \right\rangle$ in the denominator of Eq.~(4), and we will follow a similar procedure here for consistency with previous work\cite{lapillonne:2011, kumar:2021}. $k_\perp$ is averaged over the eigenmode envelope $\hat{\Phi}_{k_{x},k_{y}}(z)$, and given by

\begin{equation}
    \left\langle k_{\bot}^{2} \right\rangle=\frac{\sum_{k_{x}}^{}\int_{}^{}(g^{xx}k_{x}^{2}+2g^{xy}k_{x}k_{y}+g^{yy}k_{y}^{2})\left| \hat{\Phi}_{k_{x}k_{y}}(z) \right|^{2} J dz}{\sum_{k_{x}}^{}\int_{}^{}\left|\hat{\Phi}_{k_{x}k_{y}}(z)  \right|^{2} J dz},
\end{equation}
where $J$ is the Jacobian and $g^{xx}$, $g^{xy}$, $g^{yy}$ are geometric coefficients $g^{\mu \nu}= \nabla \mu \cdot \nabla \nu$ in the field-line following coordinates\cite{lapillonne:2011}. Note that  $\left\langle k_{\bot}^{2} \right\rangle$ given in Eq.~(5) is a function of $k_y$.

\subsection{Saturation rules}

Admittedly, though there may be validity in the quasilinear expression for the fluxes, the parameter dependence of the fluctuation amplitude is difficult to determine without running a nonlinear gyrokinetic simulation many times over a range of parameters\cite{kinsey:2015}. However, we can gain some knowledge of the transport properties by comparing different saturation rules and the sensitivity of the observed trends. In this paper, we compare three common saturation rules and give simple scaling arguments for their origin where they exist.  The first saturation rule given in Eq.~(4) can be obtained by 
balancing linear growth with the ${\bf E} \times {\bf B}$ advection. For example, the mode would saturate when $\frac{\partial \delta n}{\partial t}$ balances ${\bf v}_E \cdot \nabla \delta n$, where ${\bf v}_E$ is the ${\bf E} \times {\bf B}$ drift resulting in
\begin{equation}
\gamma \delta n_k \sim \frac{k_\perp^2}{B}  \left| \phi_k \right| \left| \delta n_k \right|, \nonumber
\end{equation}
or
\begin{equation}
\frac{e \left| \phi_k \right|}{T}  \sim \frac{e B}{T} \frac{\gamma_k}{k_\perp^2}, \nonumber
\end{equation}
which is the scaling in Eq.~(4) and commonly used\cite{fable:2010,lapillonne:2011,tirkas:2023}. The saturation rule can also be obtained from wave particle trapping resonant particles and gives the correct saturation level in slab geometry\cite{parker:1993}.

The second saturation rule we will use can come from a dimensional argument where the diffusion coefficient is simply set to $D= \gamma /  k_\perp^2$.  Then, $D \nabla n_0 = \left\langle v_{Ex} \delta n \right \rangle$ to obtain
\begin{equation}
\frac{e \left| \phi_k \right|^2}{T}  =A_0^2 \frac{e B}{T} \frac{1}{L k_y} \frac{\gamma_k}{k_\perp^2},
\end{equation}
where $L$ is the gradient scale length.  $L=L_n$, the density gradient scale length in this argument. A similar calculation could be made for the thermal diffusivity, so we write $L$ in Eq.~(6) more generally.  Eq.~(6) is similar to the saturation rule used in QuaLiKiz\cite{bourdelle:2007} and in earlier work\cite{dannert:2005}.

Finally, the third saturation rule we will use for comparison is the following
\begin{equation}
\frac{e \left| \phi_k \right|^2}{T}  = A_0^2 \frac{e B}{T}  \frac{\gamma_k}{k_\perp^2},
\end{equation}
which has a similar ${\gamma \over k_\perp^2}$ scaling as Eq.~(6), but does not diverge as $k_y$ approaches zero. The saturation rule given in Eq.~(7) has been used previously for comparison to nonlinear gyrokinetic simulation and 
experiment\cite{kumar:2021}.  Our goal here is not to develop a transport model that accurately reproduces nonlinear gyrokinetic simulation. Rather, we admit the saturation rule is the weakness in any weak turbulence model.  Therefore, we present results for the three saturation rules above, and in some sense "scan" the sensitivity of the fluxes to the saturation rule.

\subsection{Saturation levels and quasilinear fluxes}

Fig.~4 shows the three saturation rules, Eqs.~(4), (6) and (7), with $A_0=1$ and $T=T_i$, along with the TGLF SAT1 result. GENE gyroBohm units are used in which $\phi$ is normalized by ${R \over \rho_i} {e \phi \over T_i}$.  I.e., to convert to SI units [V$^2$], 
take the values presented in Fig.~4 and multiply by $\left ({\rho_i \over R} {T_i \over e} \right )^2$.
The saturation rules Eq.~(4), Eq.~(6) and Eq.~(7) are labeled ``Lapillonne(2011)'', ``Bourdelle(2007)'', and ``Kumar(2021)'', respectively, simply for the convenience of the reader. TGLF SAT1 is labeled ``SAT1''. No indication of the relative validity of the various models should be made since we are only using the various saturation rules for comparison.  Additionally, the various theory-based transports models calculate the fluxes differently than the way we do here.  Namely, we use GENE linear results. 
The overall level of each saturation rule, i.e. the value of $A_0$, has little meaning since quasilinear transport models calibrate the saturation rule using a constant coefficient. 

The saturation rules show similar trends, peaking at $k_y \rho_i\sim$ 0.2-0.3.  There is some variation in the width of the spectra. SAT1 gives the narrowest spectrum, and not surprisingly Eq.~(7) gives a broader spectrum  compared to Eq.~(4) and (6) due to the power of the ${\gamma_k \over k_\perp^2}$ term and the lack of a ${1 \over k_y}$ factor. It is interesting that Eq.~(6) and SAT1 are somewhat similar.

\begin{figure}[H]
\centering
\includegraphics[width=0.55\textwidth]{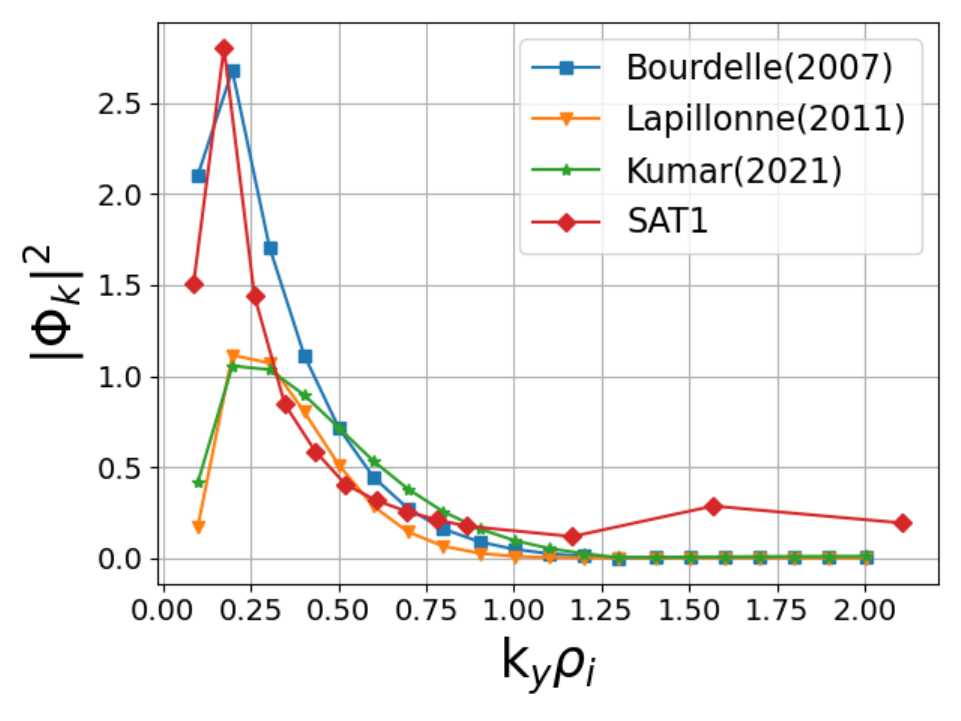}
\label{fig:fig4} 
\caption{The three saturation rules obtained from linear GENE along with TGLF SAT1 in GENE gyroBohm units described in the text.}
\end{figure}

Next, we compare the quasilinear flux obtained from the normalized GENE quasilinear flux and the three saturation rules.  
The results are shown in Fig.~5. We use GENE gyroBohm units where $Q [\mbox{SI}] = \left ({v_{th} \rho_i^2 n_e T_i \over R^2} \right) Q_{\mbox{shown}}$, $\Gamma [\mbox{SI}] = \left ({v_{th} \rho_i^2 n_e \over R^2} \right ) \Gamma_{\mbox{shown}}$.
The results in Fig.~(5) have been normalized, by adjusting $A_0$ in the three saturation rules so that the total ion heat flux, $Q_i$ matches that predicted by SAT1.
The SAT1 result shown in Fig.~5 is using the TGLF model directly. 
\begin{figure*}[h]
\centering
\includegraphics[width=0.95\textwidth]{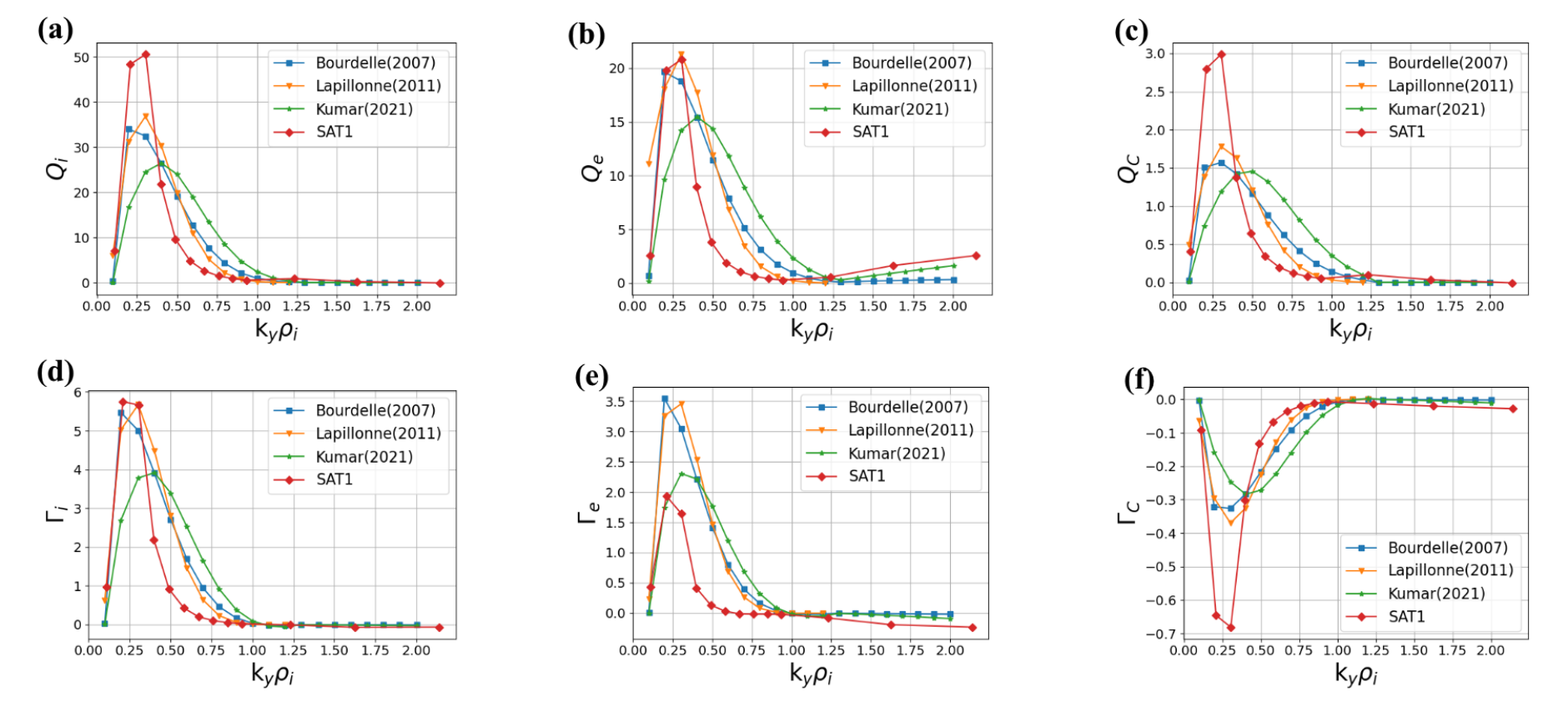}
\label{fig:fig5} 
\caption{Quasilinear fluxes from GENE versus $k_y\rho_i$ at $\rho=0.85$ for the three saturation rules in GENE gyroBohm units. (\textbf{a}) Deuterium heat flux. (\textbf{b}) Electron heat flux. (\textbf{c}) Carbon heat flux. (\textbf{d}) Deuterium particle flux. (\textbf{e}) Electron particle flux. (\textbf{f}) Carbon particle flux. Fluxes are normalized such that total flux matches SAT1.}
\end{figure*}

The Carbon flux is directed inward, as expected due to the slightly hollow profile of carbon.  The results from all four models agree qualitatively, again, with some variation in the breadth of the spectrum with TGLF showing more flux at lower $k_y$. The relatively large values of TGLF fluxes vs. $k_y$ is simply an artifact of normalizing the other quasilinear fluxes to TGLF and the fact the other models have a broader $k_y$ spectrum. TGLF shows some electron flux at higher $k_y$ that the GENE quasilinear fluxes do not.

\begin{figure*}[h]
\centering
\includegraphics[width=0.95\textwidth]{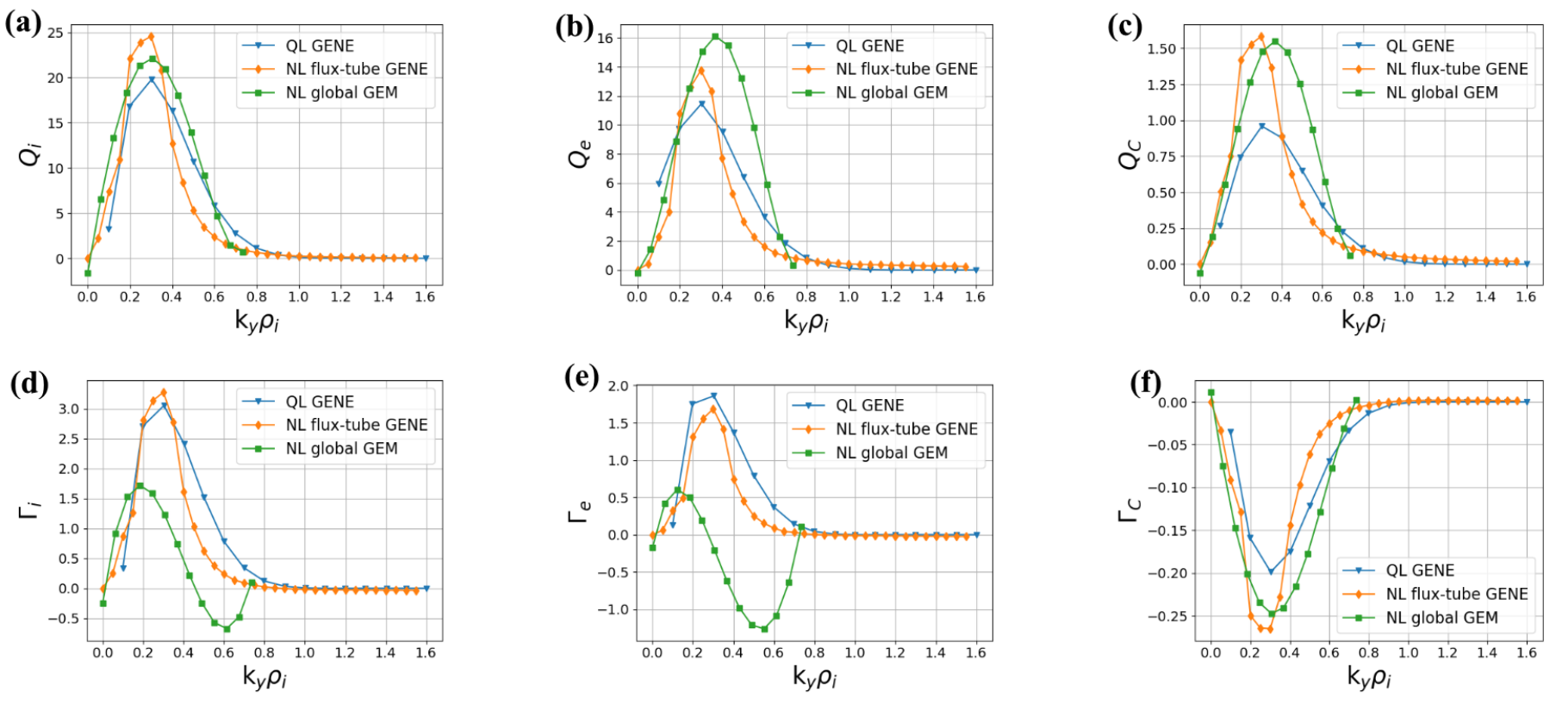}
\label{fig:fig6} 
\caption{Lapillonne QL flux model, NL from GENE, and GEM results versus $k_y\rho_i$ at $\rho=0.85$ in GENE gyroBohm units. GEM fluxes scaled by $3.49$. A) Deuterium heat flux. B) Electron heat flux. C) Carbon heat flux. D) Deuterium particle flux. E) Electron particle flux. F) Carbon particle flux.}
\end{figure*}

\section{Comparison with nonlinear gyrokinetic simulations}
\label{sec:others}

Local nonlinear flux-tube simulations were carried out using the GENE code at $\rho=0.85$ to test the validity of the quasilinear models. The perpendicular flux-tube domain used was 167$\rho_i$ (radial) $\times$ 126$\rho_i$ in size with 256 radial grid points and 32 toroidal modes with $k_y\rho_i$ ranging from $0.05$ to $1.60$. Grid resolution in the $z, v_{\parallel},$ and $ \mu$ dimensions were chosen to be $32\times32\times16$ respectively, where the values were found by running linear growth rate convergence tests. Initial runs with the value of $\beta_e = 0.85\%$ taken from Table 1 showed nonlinearly excited lower-$k_y$ micro-tearing modes (MTM) dominating the transport at earlier times, e.g. $t\frac{v_{th}}{R}\le 20$, and eventually going numerically unstable at late times. Since  electromagnetic modes were not observed in the global nonlinear GEM simulations discussed below, $\beta_e$ was reduced to 10\% of the original value, and high-quality electrostatic simulation results were obtained. This change is reasonable since the fluxes are mainly electrostatic in the  GEM simulations as nonlocal effects may help to stabilize the low-n electromagnetic modes.

Fig.~6 shows a comparison of particle and heat fluxes versus $k_y$ between local GENE, quasilinear theory and global GEM. The quasilinear fluxes shown in blue and labeled ``QL GENE'' are in good agreement with the nonlinear GENE results in orange labeled ``NL flux-tube GENE'' in Fig.~6 except the amplitude $A_0$ is normalized using the nonlinear GENE ion heat flux. This is appropriate since the value of $A_0$ is undetermined in the theory.
The global GEM results, discussed below, are scaled by a factor of $3.49$. Nonlocal effects, including profile, q, and magnetic shear variation are stabilizing. Therefore, it is typical that global calculations are more stable, hence producing lower fluxes.

The results labeled ``Nonlinear global GEM'' in Fig.~6, are nolocal nonlinear electromagnetic gyrokinetic simulations using the $\delta\!f$ particle-in-cell code GEM. For the present study of ion-scale turbulent transport, a fully drift-kinetic electron species is included using the split-weight scheme\cite{chen:2007}. To ensure a steady-state turbulence and transport, a numerical heat source is applied to all species\cite{chen:2023}, and a numerical scheme\cite{chen:2022} is used for evaluating the marker distribution which can evolve significantly in later times. The grid resolution is $(N_x,N_y,N_z)=(128,128,64)$, in the radial, binormal and parallel direction, respectively. The particle number is $32$/cell for the ion species and $64$/cell for electrons. The time step is $\Omega_p \triangle t=1$ where $\Omega_p$ is the proton gyro-frequency. The radial domain of the nonlocal simulation is $0.65<r/a<0.95$. Attempts to extend the simulation to the separatrix ($r/a=1.0$) lead to nonphysical modes near the edge. The cause of this problem is not clear, but we believe part of the reason is the uncertainty in the equilibrium configuration, including the magnetic surface shape and the density/temperature profiles. In particular, strong poloidal variation of the temperature is expected in a region of steep gradients, but such variation is not modeled in the present study since local Maxwellian distributions that vary only in the radial flux coordinate are assumed. The density and temperature gradients are reduced in a boundary layer of $\triangle r/a\sim 0.05$ near the outer domain, to avoid peaking of the turbulence near the boundary. 

In the toroidal direction, the simulation domain is a toroidal wedge which is $1/8$ of the torus, and the EM fields are filtered to include only the toroidal mode numbers $n=0, 8, 16, \ldots, 88$. We note that no nonlinear excitation of low-n electromagnetic modes, e.g. MTMs, was present in the global GEM simulations, possibly due to nonlocal stabilization, in contrast to local GENE results. Fig.~7 shows the ion energy fluxes at various radial locations from GEM. The quality of the simulation seems reasonable.
\begin{figure}[H]
\centering
\includegraphics[width=0.6\textwidth]{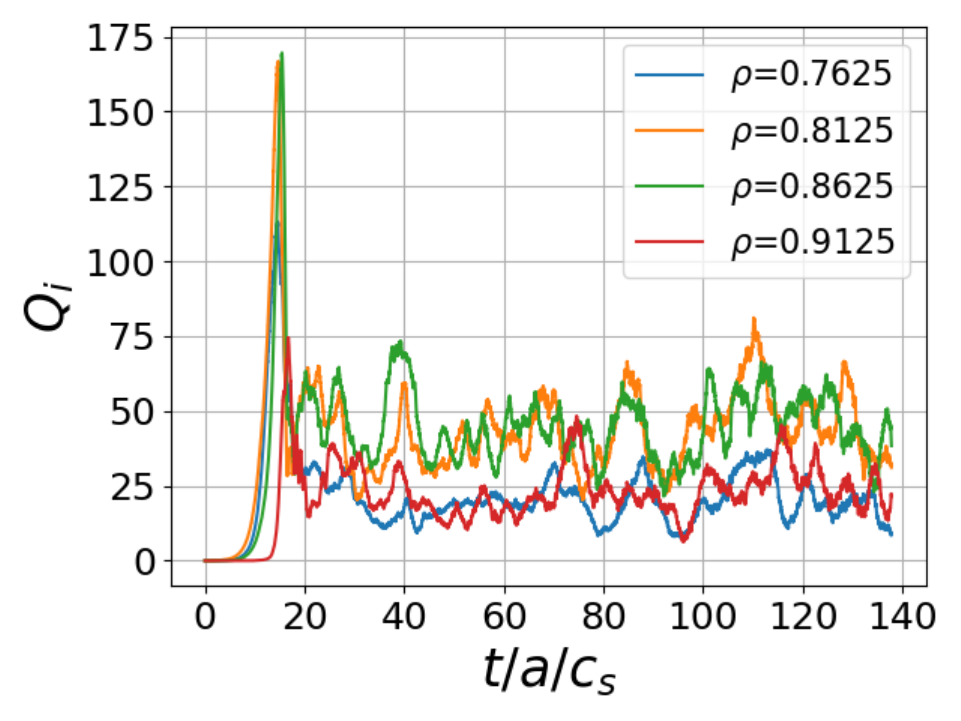}
\label{fig:fig7} 
\caption{Global GEM ion heat flux versus time in GENE gyroBohm units at multiple radial locations.}
\end{figure}

For comparison to GENE and quasilinear theory, the turbulent fluxes are decomposed into toroidal modes.
 For example, in the formula for the radial heat flux,
\begin{equation*}
    Q(r)=\frac{1}{\triangle V}\int_{\triangle V} \frac{1}{2}mv^2 \,\delta \! f \, 
\left( 
    \frac{{\bf E}\times {\bf b}}{B}
+v_\parallel \frac{\delta{\bf B}_\perp}{B} 
    \right) 
\cdot \frac{\nabla r}{|{\nabla r}|} \,d{\bf x}d{\bf v},
\end{equation*}
if ${\bf E}$ and ${\bf B}$ are replaced by a specific toroidal component, the contribution of that component to the total flux is obtained. Here $\triangle V$ is a thin toroidal annulus with a radial size of $\triangle r/a=0.025$. Results from this procedure are shown in Fig.~8. The raw results are shown as solid red triangles and the solid blue squares are a polynomial fit.  The fitted result is shown in Fig.~7. GEM is a particle code and does not evolve the distribution function spectrally in $k_y$ like the GENE calculation. Obtaining $Q(k_y)$ involves summing over particle weights for each toroidal mode in GEM and leads to statistical fluctuations in this quantity.  The smooth fit shown in Fig.~6 (solid blue squares in Fig.~8) more clearly compares the trends between models. We do the same fitting procedure for all the flux quantities in Fig.~6.

\begin{figure}[H]
\centering
\includegraphics[width=0.6\textwidth]{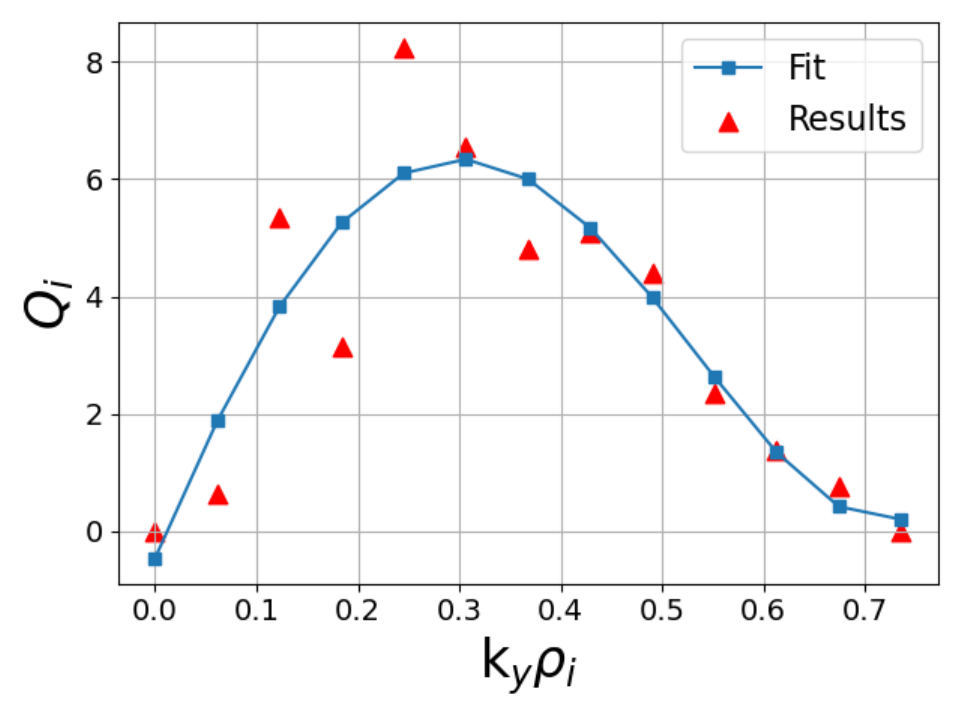}
\label{fig:fig8} 
\caption{GEM Ion heat flux versus $k_y$ and a corresponding smooth polynomial fit.}
\end{figure}


\section{Discussion and future work}

In high confinement tokamak regimes, it is reasonable to assume the quasilinear expression for the flux is valid, but there is still uncertainty in determining the fluctuation level.  Here, we compared three common saturation rules using gyrokinetic simulation to determine the quasilinear flux expression.  To our knowledge, the various common saturation rules have not been compared in this way before. Considerable realism was taken into account including experimental plasma parameters, collisionality, electromagnetics, and drift-kinetic electrons. We show that with proper normalization to nonlinear gyrokinetic simulation or experiment, the three saturation rules exhibit similar behavior.  We also compared to local and global gyrokinetic simulations.  The local nonlinear GENE result exhibited nonlinearly excited low $k_y$ electromagnetic modes that were dominant.  However, when $\beta$ was reduced in the simulation to eliminate the low $k_y$ modes, nonlinear GENE showed similar behavior as the quasilinear theory. Electromagnetic global GEM showed qualitatively similar behavior (Fig.~6).  GEM did not see a low-n electromagnetic mode, possibly due to nonlocal stabilization, e.g. variation in profiles, q, and magnetic shear in this region. Global GEM also gave a lower ion and electron particle flux. Nonlinear GENE and GEM showed a higher impurity flux. All calculations presented in this paper are quasineutral, of course. Further work will include a better understanding of the larger inward particle flux found in the nonlinear simulations.

This study only scraped the tip of the iceberg, and further investigation is needed to better understand the parametric dependence of the models for a wider variety of plasma equilibria and profiles. Even for this particular case (DIII-D 162940) it would be interesting to look at the effect of including equilibrium shear flow and the radial dependence of the fluxes.
These are topics for future work. Detailed comparisons with nonlinear simulation is challenging due to computing resources.  However, comparison between the various quasilinear models is relatively fast requiring only linear GENE calculations. Additionally, it would be useful to investigate the relative contributions of particle diffusion and convection and impurity peaking factor\cite{howard:2012,angioni:2014}. Since, in quasilinear theory, the amplitude cancels out in the determination of the peaking factor, the choice of particular saturation rule should be less important. 


\vspace{6pt} 


Work supported by the U.S. Department of Energy cooperative agreements DE-SC-000801 and DE-FG02-08ER54954. This research used resources of the National Energy Research Scientific Computing Center (NERSC), a U.S. Department of Energy Office of Science User Facility located at Lawrence Berkeley National Laboratory, operated under Contract No. DE-AC02-05CH11231.

We thank Gabriele Merlo (University of Texas, Austin) for his continued support with running the GENE code and interpreting results.
We thank Emily Belli (GA - General Atomics) and Gary Staebler (Oak Ridge National Laboratory) for access and help with TGLF and GA code modeling software. We also thank Richard Groebner (GA), Brian Grierson (GA), Shawn Haskey (Princeton Plasma Physics Laboratory), and Neeraj Kumar (General Fusion) for providing profiles, magnetic equilibria and useful discussion.

Conceptualization, project administration, supervision, funding acquisition, S. Parker; methodology, S. Parker, C. Haubrich, Q. Cai, S. Tirkas, Y. Chen; software, C. Haubrich, Q. Cai, S. Tirkas, Y. Chen; data curation, C. Haubrich; writing-original draft preparation, S. Parker, C. Haubrich, S. Tirkas, Y. Chen, writing-review and editing, S. Parker, C. Haubrich, Q. Cai, S. Tirkas, Y. Chen. All authors have read and agreed to the published version of the manuscript.

Data and code available on request.

The authors declare no conflict of interest.

\bibliographystyle{unsrt}  
\bibliography{references}

\begin{thebibliography}{10}

\bibitem{staebler:2005}
G.M. Staebler, J.E. Kinsey, and R.E. Waltz.
\newblock {{Gyro-Landau}} fluid equations for trapped and passing particles.
\newblock {\em Phys. Plasmas}, 12:102508, 2005.

\bibitem{staebler:2007}
G.M. Staebler, J.E. Kinsey, and R.E. Waltz.
\newblock A theory-based transport model with comprehensive physics.
\newblock {\em Phys. Plasmas}, 14:055909, 2007.

\bibitem{kinsey:2015}
J.~E. Kinsey, G.~M. Staebler, J.~Candy, C.~C. Petty, T.~L. Rhodes, and R.~E.
  Waltz.
\newblock {{Predictions of the near edge transport shortfall in DIII-D L-mode
  plasmas using the trapped gyro-Landau-fluid model}}.
\newblock {\em Phys. Plasmas}, 22:012507, 2015.

\bibitem{staebler:2016}
G.M. Staebler, J.~Candy, N.~T. Howard, and C.~Holland.
\newblock The role of zonal flows in the saturation of multi-scale gyrokinetic
  turbulence.
\newblock {\em Phys. Plasmas}, 23:062518, 2016.

\bibitem{kinsey:1995}
J.~Kinsey, C.~Singer, D.~Cox, and G.~Bateman.
\newblock Systematic comparison of a theory-based transport model with a
  multi-tokamak profile database.
\newblock {\em Phys. Scripta}, 52:428, 1995.

\bibitem{bateman:1998}
G.~Bateman, A.H. Kritz, J.E. Kinsey, A.J. Redd, and J.~Weiland.
\newblock Predicting temperature and density profiles in tokamaks.
\newblock {\em Phys. Plasmas}, 5:1793, 1998.

\bibitem{rafiq:2013}
T.~Rafiq, A.H. Kritz, J.~Weiland, A.Y. Pankin, and L.~Luo.
\newblock Physics basis of multi-mode anomalous transport module.
\newblock {\em Phys. Plasmas}, 20, 2013.

\bibitem{weiland:2020}
J.~Weiland, A.~Zagorodny, and T.~Rafiq.
\newblock Theory for transport in magnetized plasmas.
\newblock {\em Physica Scripta}, 95:105607, 2020.

\bibitem{bourdelle:2007}
C.~Bourdelle, X.~Garbet, F.~Imbeaux, A.~Casati, N.~Dubuit, R.~Guirlet, and
  T.~Parisot.
\newblock A new gyrokinetic quasilinear transport model applied to particle
  transport in tokamak plasmas.
\newblock {\em Physics of Plasmas}, 14(11):112501, 2007.

\bibitem{bourdelle:2016}
C.~Bourdelle, J.~Citrin, B.~Baiocchi, A.~Casati, P.~Cottier, X.~Garbet,
  F.~Imbeaux, et~al.
\newblock Core turbulent transport in tokamak plasmas: bridging theory and
  experiment with {{QuaLiKiz}}.
\newblock {\em Plasma Phys. Control. Fusion}, 58:014036, 2016.

\bibitem{citrin:2017}
J.~Citrin, C.~Bourdelle, F.~J. Casson, C.~Angioni, N.~Bonanomi, Y.~Camenen,
  X.~Garbet, L.~Garzotti, T.~Görler, O.~Gürcan, F.~Koechl, F.~Imbeaux,
  O.~Linder, K.~van~de Plassche, P.~Strand, G.~Szepesi, and JET Contributors.
\newblock Tractable flux-driven temperature, density, and rotation profile
  evolution with the quasilinear gyrokinetic transport model qualikiz.
\newblock {\em Plasma Phys. Control. Fusion}, 59(12):124005, 2017.

\bibitem{stephens:2021}
C.D. Stephens, X.~Garbet, J.~Citrin, C.~Bourdelle, K.L. van~de Plassche, and
  F.~Jenko.
\newblock Quasilinear gyrokinetic theory: a derivation of {QuaLiKiz}.
\newblock {\em J. Plasma Phys.}, 87(4), 2021.

\bibitem{dannert:2005}
T.~Dannert and F.~Jenko.
\newblock Gyrokinetic simulation of collisionless trapped-electron mode
  turbulence.
\newblock {\em Phys. Plasmas}, 12(7):072309, 2005.

\bibitem{fable:2010}
E.~Fable, C.~Angioni, and O.~Sauter.
\newblock The role of ion and electron electrostatic turbulence in
  characterizing stationary particle transport in the core of tokamak plasmas.
\newblock {\em Plasma Phys. Control. Fusion}, 52:015007, 2010.

\bibitem{lapillonne:2011}
X~Lapillonne, S~Brunner, O~Sauter, L~Villard, E~Fable, T~G{\"o}rler, F~Jenko,
  and F~Merz.
\newblock Non-linear gyrokinetic simulations of microturbulence in {TCV}
  electron internal transport barriers.
\newblock {\em Plasma Physics and Controlled Fusion}, 53(5):054011, apr 2011.

\bibitem{tirkas:2023}
S.~Tirkas, H.~Chen, G.~Merlo, F.~Jenko, and S.~Parker.
\newblock Zonal flow excitation in electron-scale tokamak turbulence.
\newblock {\em Nucl. Fusion}, 63(2):026015, 2023.

\bibitem{angioni:2009c}
C.~Angioni, E.~Fable, M.~Greenwald, M.~Maslov, A.G. Peeters, H.~Takenaga, and
  H.~Weisen.
\newblock Particle transport in tokamak plasmas, theory and experiment.
\newblock {\em Plasma Phys. Control. Fusion}, 51:124017, 2009.

\bibitem{howard:2021b}
N.T. Howard, C.~Holland, T.L. Rhodes, J.~Candy, P.~Rodriguez-Fernandez,
  M.~Greenwald, A.E. White, and F.~Sciortino.
\newblock The role of ion and electron-scale turbulence is setting heat and
  particle transport in the {{DIII-D}} {{ITER}} baseline scenario.
\newblock {\em Phys. Plasmas}, 28:072502, 2021.

\bibitem{loarte:2020}
A.~Loarte.
\newblock Required r \& d in existing fusion facilities to support the iter
  research plan.
\newblock Technical Report ITR-20-008, ITER Organization, 2020.

\bibitem{angioni:2017}
C.~Angioni, R.~Bilato, F.J. Casson, E.~Fable, P.~Mantica, T.~Odstrcil,
  M.~Valisa, {{ASDEX Upgrade Team}}, and {{JET Contributors}}.
\newblock Gyrokinetic study of turbulent convection of heavy impurities in
  tokamak plasmas at comparable ion and electron heat fluxes.
\newblock {\em Nucl. Fusion}, 57:022009, 2017.

\bibitem{kumar:2021}
N.~Kumar, Y.~Camenen, S.~Benkadda, C.~Bourdelle, A.~Loarte, A.R. Polevoi,
  F.~Widmer, and JET contributors.
\newblock Turbulent transport driven by kinetic ballooning modes in the inner
  core of jet hybrid h-modes.
\newblock {\em Nuclear Fusion}, 61(3):036005, jan 2021.

\bibitem{staebler:2021}
G.M. Staebler, E.~Belli, J.~Candy, J.E. Kinsey, H.~Dudding, and B.~Patel.
\newblock Verification of a quasi-linear model for gyrokinetic turbulent
  transport.
\newblock {\em Nucl. Fusion}, 61:116007, 2021.

\bibitem{candy:2016}
J.~Candy, E.A. Belli, and R.V. Bravenec.
\newblock A high-accuracy {{Eulerian}} gyrokinetic solver for collisional
  plasmas.
\newblock {\em J. Comput. Phys.}, 324:73, 2016.

\bibitem{jenko:2000b}
F.~Jenko, W.~Dorland, M.~Kotschenreuther, and B.N. Rogers.
\newblock Electron temperature gradient driven turbulence.
\newblock {\em Phys. Plasmas}, 7:1904, 2000.

\bibitem{gorler:2011}
T.~G{\"o}rler, X.~Lapillonne, S.~Brunner, T.~Dannert, F.~Jenko, F.~Merz, and
  D.~Told.
\newblock {{The global version of the gyrokinetic turbulence code GENE}}.
\newblock {\em J. Comput. Phys.}, 230:7053, 2011.

\bibitem{chen:2003b}
Y.~Chen and S.E. Parker.
\newblock A $\delta\!f$ particle method for gyrokinetic simulations with
  kinetic electrons and electromagnetic perturbations.
\newblock {\em J. Comput. Phys.}, 189:463, 2003.

\bibitem{chen:2007}
Y.~Chen and S.~E. Parker.
\newblock Electromagnetic gyrokinetic delta-f particle-in-cell turbulence
  simulation with realistic equilibrium profiles and geometry.
\newblock {\em J. Comput. Phys.}, 220:839, 2007.

\bibitem{hassan:2022}
E.~Hassan, D.R. Hatch, M.R. Halfmoon, M.~Curie, M.T. Kotchenreuther, S.M.
  Mahajan, G.~Merlo, R.J. Groebner, A.O. Nelson, and A.~Diallo.
\newblock Identifying the microtearing modes in the pedestal of diii-d h-modes
  using gyrokinetic simulations.
\newblock {\em Nucl. Fusion}, 62(2):026008, 2021.

\bibitem{hassan:2021}
E.~Hassan, D.R. Hatch, W.~Guttenfelder, Y.~Chen, and S.~Parker.
\newblock Gyrokinetic benchmark of the electron temperature-gradient
  instability in the pedestal region.
\newblock {\em Phys. Plasmas}, 28(6):062505, 2021.

\bibitem{curie:2022}
M.T. Curie, J.L. Larakers, D.R. Hatch, A.O. Nelson, A.~Diallo, E.~Hassan,
  W.~Guttenfelder, M.~Halfmoon, M.~Kotschenreuther, R.D. Hazeltine, S.M.
  Mahajan, R.J. Groebner, J.~Chen, DIII-D Team, C.~Perez von Thun,
  L.~Frassinetti, S.~Saarelma, C.~Giroud, JET Contributors, and M.M. Tennery.
\newblock A survey of pedestal magnetic fluctuations using gyrokinetics and a
  global reduced model for microtearing stability.
\newblock {\em Phys. Plasmas}, 29(4):042503, 2022.

\bibitem{miller:1998}
R.L. Miller, M.S. Chu, J.M. Greene, Y.R. Lin-Liu, and R.E. Waltz.
\newblock Noncircular, finite aspect ratio, local equilibrium model.
\newblock {\em Phys. Plasmas}, 5:973, 1998.

\bibitem{gorler:2016}
T.~Görler, N.~Tronko, W.~A. Hornsby, A.~Bottino, R.~Kleiber, C.~Norscini,
  V.~Grandgirard, F.~Jenko, and E.~Sonnendrücker.
\newblock {Intercode comparison of gyrokinetic global electromagnetic modes}.
\newblock {\em Physics of Plasmas}, 23(7):072503, 07 2016.

\bibitem{parker:1993}
S.~E. Parker, W.~Dorland, R.~A. Santoro, M.~A. Beer, Q.~P. Liu, W.~W. Lee, and
  G.~W. Hammett.
\newblock {Comparisons of gyrofluid and gyrokinetic simulations*}.
\newblock {\em Physics of Plasmas}, 1(5):1461--1468, 05 1994.

\bibitem{chen:2023}
Y.~Chen, J.~Cheng, and S.~E. Parker.
\newblock {Effects of the heat source on the steady-state transport in
  gradient-driven global gyrokinetic simulations}.
\newblock {\em Physics of Plasmas}, 30(1):014502, 01 2023.

\bibitem{chen:2022}
Y.~Chen, J.~Cheng, and S.~E. Parker.
\newblock Evolution of the marker distribution in gyrokinetic $\delta\!f$
  particle-in-cell simulations.
\newblock {\em Plasma Phys.}, 29:073901, 2022.

\bibitem{howard:2012}
N.T. Howard, M.~Greenwald, D.R. Mikkelsen, M.L. Reinke, A.E. White, D.~Ernst,
  Y.~Podpaly, and J.~Candy.
\newblock Quantitative comparison of experimental impurity transport with
  nonlinear gyrokinetic simulation in an {{Alcator C-Mod L-mode}} plasma.
\newblock {\em Nucl. Fusion}, 52:063002, 2012.

\bibitem{angioni:2014}
C.~Angioni, P.~Mantica, T.~P\"utterich, M.~Valisa, M.~Baruzzo, E.A. Belli,
  P.~Belo, F.J. Casson, C.~Challis, P.~Drewelow, C.~Giroud, N.~Hawkes, T.C.
  Hender, J.~Hobirk, T.~Koskela, L.~Lauro Taroni, C.F. Maggi, J.~Mlynar,
  T.~Odstrcil, M.L. Reinke, M.~Romanelli, and {{JET EFDA Contributors}}.
\newblock Tungsten transport in {{JET H-mode}} plasmas in hybrid scenario,
  experimental observations and modelling.
\newblock {\em Nucl. Fusion}, 54:083028, 2014.

\end{thebibliography}

\end{document}